\documentclass[10pt]{article}
\usepackage{cite}
\usepackage{graphicx,subfigure}
\usepackage{psfrag}
\usepackage{url}
\usepackage{algorithm,algorithmic}
\usepackage{array}
\usepackage{amsmath,amssymb}
\usepackage{wasysym}
\usepackage{paralist}

\newtheorem{theorem}{Theorem}

\begin{document}
\title{NP-completeness Proof: RBCDN Reduction Problem}
\author{Sujogya Banerjee, Shahrzad  Shirazipourazad, Pavel Ghosh and Arunabha Sen \\
\small Computer Science and Engineering Program\\
\small School of Computing, Informatics and Decision System Engineering\\
\small Arizona State University\\
\small Tempe, Arizona 85287\\
\small Email: \{sujogya, sshiraz1, pavel.ghosh, asen \}@asu.edu
}

\maketitle
Suppose $\lbrace R_1, \ldots, R_k\rbrace$ is the set of all possible regions \cite{Sen06} of graph $G$. Consider a $k$-dimensional vector $C$ whose $i$-th entry, $C[i]$, indicates the number of connected components in which $G$ decomposes when all nodes in $R_{i}$ fails. Then, {\em region-based component decomposition number (RBCDN)} of graph $G$ with region $R$ is defined as $\alpha_R(G) = \max_{1 \leq i \leq k} C[i]$.

 Suppose the {\em RBCDN} of $G$ with region $R$ is $\alpha_{R}(G)$. If $\alpha_{R}(G)$ is considered to be too high for the application and it requires {\em RBCDN} of the network not to exceed $\alpha_{R}(G) - K$, for some integer $K$. Assuming each additional link $l_{i}$ that can be added to the network has a cost $c(i)$ associated with it, find the least cost link augmentation to the network so that its {\em RBCDN} is reduced from $\alpha_{R}(G)$ to $\alpha_{R}(G) - K$. Formal description of the decision version of this problem is given below.

\vspace{0.05in}
\noindent
{\tt \emph{RBCDN} Reduction Problem (\emph{RBCDN-RP})}\\
INSTANCE: Given\\
(i) a graph $G = (V,E)$ where $V= \{v_1, \ldots , v_n \}$ and $E = \{e_1, \ldots , e_m\}$ are the sets of nodes and links respectively,\\
(ii) the layout of $G$ on a two dimensional plane $LG = (P,L)$ where $P = \{p_1, \ldots , p_n\}$ and $L = \{l_1, \ldots , l_m\}$ are the sets of points and lines on the 2-dimensional plane,\\
(iii) region $R$   defined to be a circular area of radius $r$. \\
(iv) cost function $c(e) \in \mathbb{Z}^+$, $\forall e \in {\bar E}$, where ${\bar E}$ is complement of the link set $E$.\\
(v) integers $C$ and $K$ ($K \le \alpha_{R}(G)$, where $\alpha_{R} (G)$ is the \emph{RBCDN} of $G$).

\vspace{0.05in}
\noindent
QUESTION: Is it possible to reduce the \emph{RBCDN} of $G$ by $K$ by adding edges to $G$ (from the set ${\bar E}$) so that the total cost of the added links is at most $C$?

\section{NP-Completeness Proof of \emph{RBCDN-RP}}
\label{sec:robustDesignNP}
\vspace{0.05in}
\noindent
We prove that \emph{RBCDN-RP} is NP-complete by a transformation from the {\tt Hamiltonian Cycle in Planar Graph Problem  (HCPGP)} which is known to be NP-complete \cite{GarJon}. A {\em Hamiltonian Cycle} in an undirected graph $G = (V, E)$ is a simple cycle that includes all the nodes. A graph is a {\em planar} if it can be embedded in a plane by mapping each node to a unique point in the plane and each edge is a line connecting its endpoints, so that no two lines meet except at a common endpoint \cite{GarJon}.

\vspace{0.05in}
\noindent
{\tt Hamiltonian Cycle in Planar Graph Problem  (HCPGP)}\\
INSTANCE: Given an undirected planar graph $G = (V, E)$.

\vspace{0.05in}
\noindent
QUESTION: Does $G$ contains a Hamiltonian Cycle?

\vspace{0.05in}
\begin{theorem}
\emph{RBCDN-RP}  is NP-complete.
\end{theorem}

\vspace{0.05in}
\noindent
\emph{Proof:} It is easy to verify whether a set of additional edges of total cost $\leq C$ reduces the \emph{RBCDN} of graph $G$ with region $R$ from $\alpha_R(G)$ to $\alpha_R(G) - K$. Therefore \emph{RBCDN-RP} is in NP.

\begin{figure}[!t]
\centering
\includegraphics[width=0.7\textwidth, keepaspectratio]{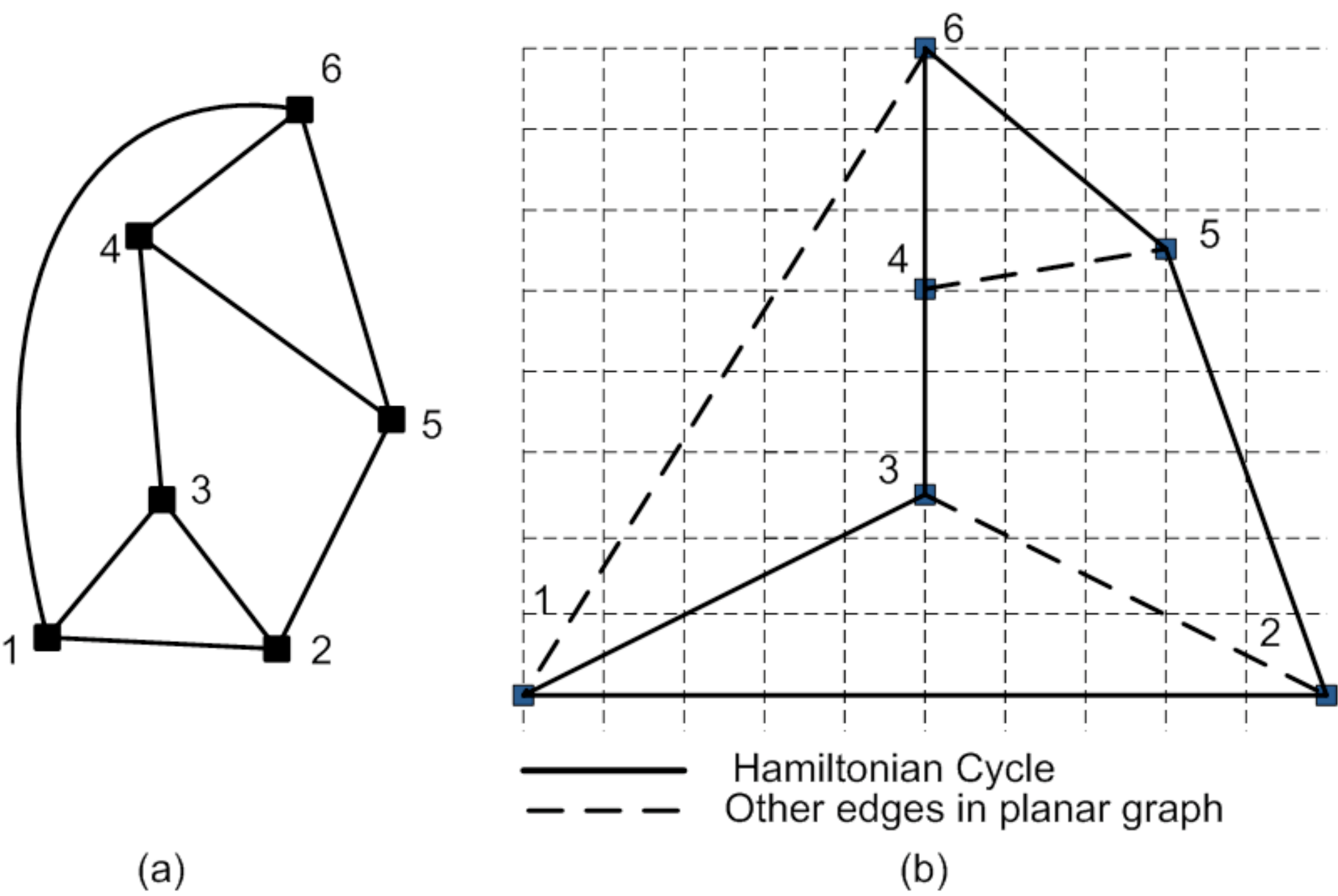}
\vspace{-0.1in}
\caption{Transformation of a \emph{HCPGP} instance to a \emph{RBCDN-RP} instance}
\label{design}
\end{figure}
From an instance of the \emph{HCPGP} (a planar graph $G = (V, E)$) we create an instance of the \emph{RBCDN-RP} (the layout $LG^\prime = (P^\prime, L^\prime)$ of a graph $G^\prime = (V^\prime, E^\prime)$) in the following way. First, we do a straight line embedding of the planar graph $G$ on a plane so that lines corresponding to links in $G$ do not intersect each other. Such an embedding can be carried out in polynomial time \cite{Fray90draw}. We call this layout $LG^{\prime\prime} = (P^{\prime\prime}, L^{\prime\prime})$. We create another layout $LG^\prime =  (P^{\prime\prime}, \emptyset)$, by setting
$P^{\prime} = P^{\prime \prime}$ and $L^{\prime} = \emptyset$. The graph $G^\prime$ corresponding to the layout $LG^\prime = (P^\prime, L^\prime)$ is the instance of \emph{RBCDN-RP} created from the instance of \emph{HCPGP}. We define a region $R$ to be circular area of sufficiently small radius $r$, such that if a region fails, it can only destroy (i) a single node with all links incident on it, or (ii) a single link. Since the created instance of \emph{RBCDN-RP} has no links ($E^\prime = L^\prime = \emptyset$), the \emph{RBCDN} of $G^\prime$ with region $R$ is $n$ where $n = |V^\prime| = |P^\prime|$. We set the parameters $C$ and $K$ of the instance of the \emph{RBCDN-RP} to be equal to $n$ and $n-1$ respectively. We assign costs to the links of $\bar{E^\prime}$ in the following way. The cost of a link $c(e)$ = 1, if $e \in (E \cap \bar {E^\prime})$ and $c(e)$ = $\infty$, if $e \in (\bar E \cap \bar {E^\prime})$.

If the instance of the \emph{HCPGP} has a Hamiltonian Cycle, we can use the set of links that make up the cycle, to augment the link set $E'$ of the instance $G' = (V', E')$ of the \emph{RBCDN-RP}. The augmented $G'$, ($G_{aug}'$), is now a simple cycle that involves all the nodes. With the given definition of region $R$ (a small circle of radius $r$), only one node can be destroyed when a region fails. Accordingly \emph{RBCDN} of $G_{aug}'$ is 1. It may be recalled that \emph{RBCDN} of $G'$ is $n$. Accordingly, augmentation of the link set of $G'$ reduced its \emph{RBCDN} by $n - 1$. Due to the specific cost assignment rule of the links, the total cost of link augmentation is $n$. Therefore, if the \emph{HCPGP} instance has a Hamiltonian Cycle, the \emph{RBCDN} of the instance of \emph{RBCDN-RP} can be reduced by $K$ with augmentation cost $\leq C$.

Suppose that it is possible to reduce the \emph{RBCDN} of the instance of \emph{RBCDN-RP} by $K$ with augmentation cost being at most $C$. This implies that the \emph{RBCDN} of $G^{\prime}$ can be reduced from $n$ to 1 (as $K = n - 1$) when it is augmented with additional links with total cost at most $n$ (as $C = n$). In order for the \emph{RBCDN} of $G_{aug}^\prime$ to be 1, the node connectivity of $G_{aug}^\prime$ must be at least 2. A $n$ node graph that has the fewest number of links and yet is 2-connected, is a cycle that includes all the nodes. As $G'$ had no links, this implies at least $n$ links must have been added to create the augmented graph $G_{aug}^\prime$. Given that the cost of a link $c(e)$ = 1, if $e \in (E \cap \bar {E^{\prime}})$ and $c(e)$ = $\infty$, if $e \in (\bar E \cap \bar {E^{\prime}})$, and total cost of link augmentation is at most $n$, it is clear that the links used in augmenting $G$ must be from the set $(E \cap \bar {E^{\prime}})$. These links are part of the edge set of the instance of \emph{HCPGP}. Accordingly, the instance of \emph{HCPGP} must have a Hamiltonian Cycle.

\bibliographystyle{splncs03}
\bibliography{references}
\end{document}